\documentclass{ws-mpla}
\usepackage[super]{cite}
\usepackage{graphicx}

\usepackage{amssymb}
\usepackage{amsmath,amscd}
\usepackage{color}
\usepackage{mathrsfs}
\usepackage{calligra}

\newcommand{\bcen}{\begin{center}}
\newcommand{\ecen}{\end{center}}
\newcommand{\bflr}{\begin{flushright}}
\newcommand{\eflr}{\end{flushright}}
\newcommand{\bfll}{\begin{flushleft}}
\newcommand{\efll}{\end{flushleft}}
\newcommand{\beq}{\begin{equation}}
\newcommand{\eeq}{\end{equation}}
\newcommand{\beqa}{\begin{eqnarray}}
\newcommand{\eeqa}{\end{eqnarray}}
\newcommand{\bite}{\begin{itemize}}
\newcommand{\eite}{\end{itemize}}
\newcommand{\benu}{\begin{enumerate}}
\newcommand{\eenu}{\end{enumerate}}

\newcommand{\h}{{\tilde h}}

\newcommand{\cO}{{\cal O}}
\newcommand{\eq}[1]{(\ref{#1})}
\newcommand{\p}{\partial}
\newcommand{\A}{{\cal A}}

\begin{document}

\markboth{Yugo Abe, Takeo Inami, Keisuke Izumi}
{Perturbative $S$-matrix unitarity ($S^{\dagger}S=1$) in $R_{\mu \nu} ^2$ gravity}

\catchline{}{}{}{}{}

\title{Perturbative $S$-matrix unitarity ($S^{\dagger}S=1$) in $R_{\mu \nu} ^2$ gravity}

\author{Yugo Abe}

\address{National Institute of Technology, Miyakonojo College, Miyakonojo 885-8567, Japan}

\author{Takeo Inami}

\address{Theoretical Research Division, Nishina Center, RIKEN, Wako 351-0198, Japan\\
Institute of Physics, VAST, Bo Ho, Hanoi, Vietnam}

\author{Keisuke Izumi}

\address{Kobayashi-Maskawa Institute, Nagoya University, Nagoya 464-8602, Japan\\
Department of Mathematics, Nagoya University, Nagoya 464-8602, Japan}

\maketitle


\begin{abstract}
We show that in the quadratic curvature theory of gravity, or simply $R_{\mu \nu} ^2$ gravity, the tree-level unitariy bound  (tree unitarity) is violated in the UV region  but an analog for $S$-matrix unitarity ($SS^{\dagger} = 1$) is satisfied. 
This theory is renormalizable, and hence the failure of tree unitarity is a counter example of Llewellyn Smith's conjecture on the relation between them. We have recently proposed a new conjecture that $S$-matrix unitarity gives the same conditions as renormalizability. We verify that $S$-matrix unitarity holds in the matter-graviton scattering at tree level in the $R_{\mu \nu} ^2$ gravity, demonstrating our new conjecture.

\end{abstract}


\section{Introduction} 
Unitarity and renormalizablity  are the two key conditions when we wish to quantize a gravitational theory. 
The unitarity bound is a necessary condition for unitarity, 
and  it has had a few important consequences%
\footnote{
The unitarity bound gives the upper bound on Higgs-boson mass\cite{Lee:1977eg}, and
an interesting connection has recently been derived\cite{Nagai:2014cua,Nagai:2019tgi} between perturbative unitarity 
constraints on $S$-matrix and finiteness (rather than renormalizability) of 
physical quantities in 
Higgs field theories.
}.
The unitarity bound  is most often used in the lowest order (tree level) of the perturbation. 
It is called {\it tree unitarity}. 
The most remarkable thing of tree unitarity is that
it already gives the information  
on the perturbative UV completion to all orders, 
as shown by Cornwall et al.\cite{CLT}.
In particular, it tells that $2\to2$ scattering amplitudes $\A(2\to2)$ should obey the high energy behavior
\begin{eqnarray}
\A(2\to2) \sim E^n, \qquad n \le 0, \qquad \mbox{as$\quad E\to\infty$}.  
\label{Aineq}
\end{eqnarray}

Llewellyn Smith gave a conjecture on the relation between renormalizablity and tree unitarity in the UV region
 in quantum field theories (QFT) \cite{Bell,L.Smith,CLT}.
The two notions are related to each other as necessary conditions for the perturbative UV completion. 
This conjecture has had a few strong supports:
gauge theories \cite{Lee:1977eg,Bell,L.Smith,CLT} and Lifshitz-type theories \cite{FIIK1,FIIK2}.
Remarkably Einstein gravity is also an example consistent with this conjecture; 
it is a non-renormalizable theory and tree unitarity \eq{Aineq} is violated in elastic graviton scattering\cite{DeWitt3,BG}.  
This conjecture is very powerful in the sense that the information of the 
UV completion can be obtained by a simple tree-level calculation of the unitarity bound.

Unfortunately, we have a counter-example of this conjecture, quadratic curvature gravity or simply $R_{\mu \nu} ^2$ gravity.
The theory is a renormalizable theory \cite{Stelle}, but the unitarity is violated due to the negative norm states%
\footnote{
Renormalizability is achieved due to the negative norm states, which are essentially the same as Pauli-Villars fields, 
and thus 
one may think that 
it is a deception. 
Nevertheless, theories with higher order derivatives have been actively studied in  both QFT and phenomenology
\cite{Lee:1969fy,Lee:1970iw,Cutkosky:1969fq,Nakanishi,Grinstein:2007mp,Cheng:2001du,Ohta:2016jvw}. 
In this letter, we do not go into the problem of negative norm, but 
investigate the relation between $S$-matrix unitarity and the finiteness of the number of conter terms.
}. 
We have recently proposed an alternative conjecture applicable to theories with negative norm states.
It is crucial to realize that 
unitarity is composed of two properties: $i)$ norm positivity and $ii)$ identity $SS^{\dagger} = 1$. 
We call the latter $S$-matrix unitarity.%
\footnote{$SS^{\dagger} = 1$ in pseudo norm space is called the pseudo unitarity. 
Here, to emphasize that our conjecture is applicable in either norm space or pseudo norm space, 
we use the term ``$S$-matrix unitarity''.}
Our conjecture is that an analog of tree unitarity for $S$-matrix unitarity ($SS^{\dagger} = 1$) in the UV region
 is related to renormalizability\cite{Abe:2018rwb},
that is to say,  in Llewellyn Smith's conjecture 
only the $S$-matrix unitarity is responsible for the renormalizability.
Our conjecture allows us to investigate the renormalizablity 
in theories with negative norm states, 
by a simple tree-level calculation of the $S$-matrix unitarity. 
To verify the conjecture, we will study the role of $SS^{\dagger} = 1$ in graviton scattering in $R_{\mu\nu}^2$ gravity
and  have studied scalar field theories with higher order derivatives\cite{Abe:2018rwb}.

In this letter
we confirm that our conjecture holds true in $R_{\mu \nu} ^2$ gravity by the calculation of matter-graviton two-body scattering. 
$S$-matrix unitarity ($SS^{\dagger} = 1$) gives the optical theorem,
\beq
2\mbox{Im}\, T(\Psi \rightarrow \Psi) = \Sigma_{\Phi} \epsilon_{\Phi} |T(\Psi \rightarrow \Phi)|^2 , 
\label{OT}
\eeq
where $\epsilon_{\Phi} = +1 (-1)$ (using an adequate normalization) if the norm of $\Phi$ is positive (negative) 
and the sum is taken over all intermediate normalized on-shell states $\Phi$.
The bound of the absolute value of the left hand side in Eq.\eq{OT} on $|T|$ gives the unitarity bound in cases without negative norm states, {\it i.e.} all $\epsilon_{\Phi}$'s are $+1$, 
\beq
2\left| T(\Psi \rightarrow \Psi) \right| \ge \Sigma_{\Phi} |T(\Psi \rightarrow \Phi)|^2 , 
\label{OTineq}
\eeq
instead of the equality \eq{OT}.

Tree unitarity is the  unitarity bound \eq{OTineq} in which both amplitudes $T(\Psi\to\Psi)$ and $T(\Psi\to\Phi)$ are computed at tree level.
Although the leading contribution at tree level in each side of the inequality \eq{OTineq} has different power of the coupling constant, 
the difference allows us to compare the scattering amplitudes among the different powers and then 
tree unitarity provides an evaluation of whether the theory has the perturbative UV completion\cite{CLT}. 
What we evaluate in this letter is an analog of the unitarity bound \eq{OTineq} for
$S$-matrix unitarity 
 in  $R_{\mu \nu} ^2$ gravity,  
\beq
2\left| T(\Psi \rightarrow \Psi) \right| \ge \Sigma_{\Phi} \epsilon_{\Phi} |T(\Psi \rightarrow \Phi)|^2  
\label{SOTineq}
\eeq
for tree level amplitudes of scalar-graviton two-body scattering in the UV region.
Hence, we will discuss the case with $\Psi$ and $\Phi$ decribing one-scalar-one-graviton states.
The details of the computation of the amplitudes will be reported in the forthcoming paper\cite{AII}.

%
\section{$ R_{\mu \nu} ^2$ gravity coupled with a scalar field $\phi$}
We begin by writing the {\it renormalizable} action $S=S_{gravity}+S_{matter}$ of $R_{\mu \nu} ^2$ gravity coupled with a matter scalar field $\phi$ \cite{Elizalde:1994gv},
\beqa
&&S_{gravity}=\int d^4x \sqrt{-g}\left( \Lambda + \frac{1}{\kappa^2} R + \alpha R^2 + \beta R_{\mu \nu} ^2 \right), \label{actiong}\\
&&S_{matter}=\int d^4x \sqrt{-g}\left(-\frac12 g^{\mu \nu} \partial_{\mu}\phi\partial_{\nu}\phi - \frac12 m^2 \phi^2
-\frac{1}{4!} \lambda \phi^4 + \xi \phi^2 R 
\right).
\label{actionm}
\eeqa
We consider the tree-level amplitude of the matter-graviton scattering in the flat spacetime, where the assumption of $\Lambda = 0$ is required 
and the $\phi^4$ term does not contribute to matter-graviton 
scattering.
Graviton field $h_{\mu \nu} (= g_{\mu \nu} -\eta_{\mu \nu})$ contains massless field $H_{\mu \nu}$ and massive one $I_{\mu \nu}$, 
\beq 
h_{\mu \nu} = H_{\mu \nu} + I_{\mu \nu} . 
\eeq
$H_{\mu \nu}$ is composed of a positive-norm massless spin-2 field with 2 degrees of freedom (DOF) denoted by $H^{(\sigma)}$ with 
$(\sigma)= (2,e), (2,o)$, while the $I_{\mu \nu}$ is composed of a negative-norm massive spin-2 field with 5 DOF denoted by $I^{(\tau)}$ with $(\tau)$ being $(2,e)$, $(2,o)$, $(1,e)$, $(1,o)$, $(0)$, and a positive-norm scalar (spin-0) graviton $I^{(S)}$. The precise meaning of the suffix $(\sigma)$ and $(\tau)$ will be explained later. The fields $H_{\mu \nu}$ and $I_{\mu \nu}$ are expanded in terms of spin-polarization bases $e_{\mu \nu}^{(\sigma)}, e_{\mu \nu}^{(\tau)}$ and $\theta_{\mu\nu}/\sqrt{3}$. In momentum space, they are written as
\beqa
&&H_{\mu \nu}(p)= \sum_{\sigma} H^{(\sigma)}(p) e_{\mu \nu}^{(\sigma)}(p), \\
&&I_{\mu \nu}(p)= \sum_{\tau} I^{(\tau)}(p) e_{\mu \nu}^{(\tau)}(p)+ \frac1{\sqrt{3}} I^{(S)}(p) \theta_{\mu\nu} .
\eeqa
where 
summations of $\sigma$ and $\tau$ are done with respect to 2 massless and 5 massive spin-2 degrees of fredom respectively.
Concrete forms of  $e_{\mu \nu}^{(\sigma)}(p)$, $e_{\mu \nu}^{(\tau)}(p)$ and $\theta_{\mu\nu}$ will be given later.

\begin{figure}[tb]
  \begin{center}
    \includegraphics[clip,width=3.0cm]{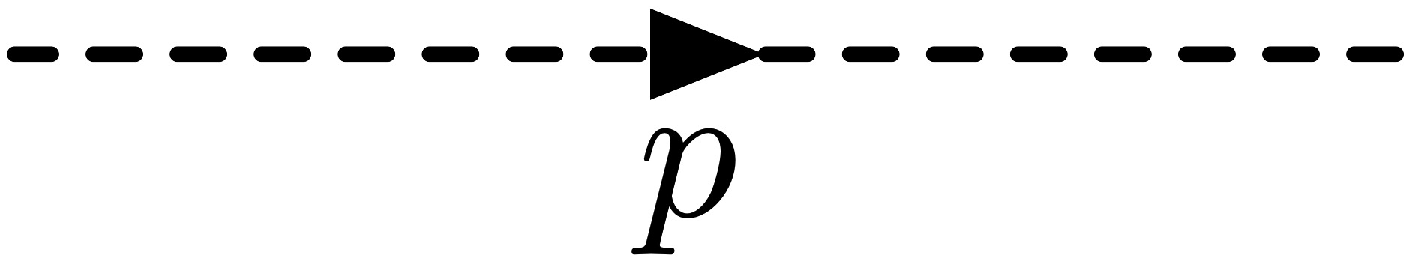}
    \hspace{5mm}
    \includegraphics[clip,width=3.0cm]{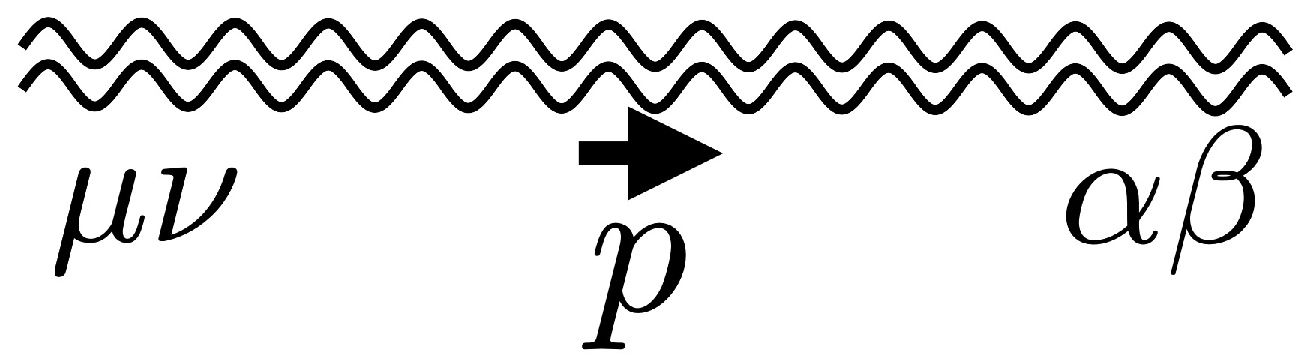}
    \caption{Scalar (left) and graviton (right) propagators.}
   \label{Fig:pro}
  \end{center}
\end{figure}

\begin{figure}[tb]
  \begin{center}
    \includegraphics[clip,width=3.6cm]{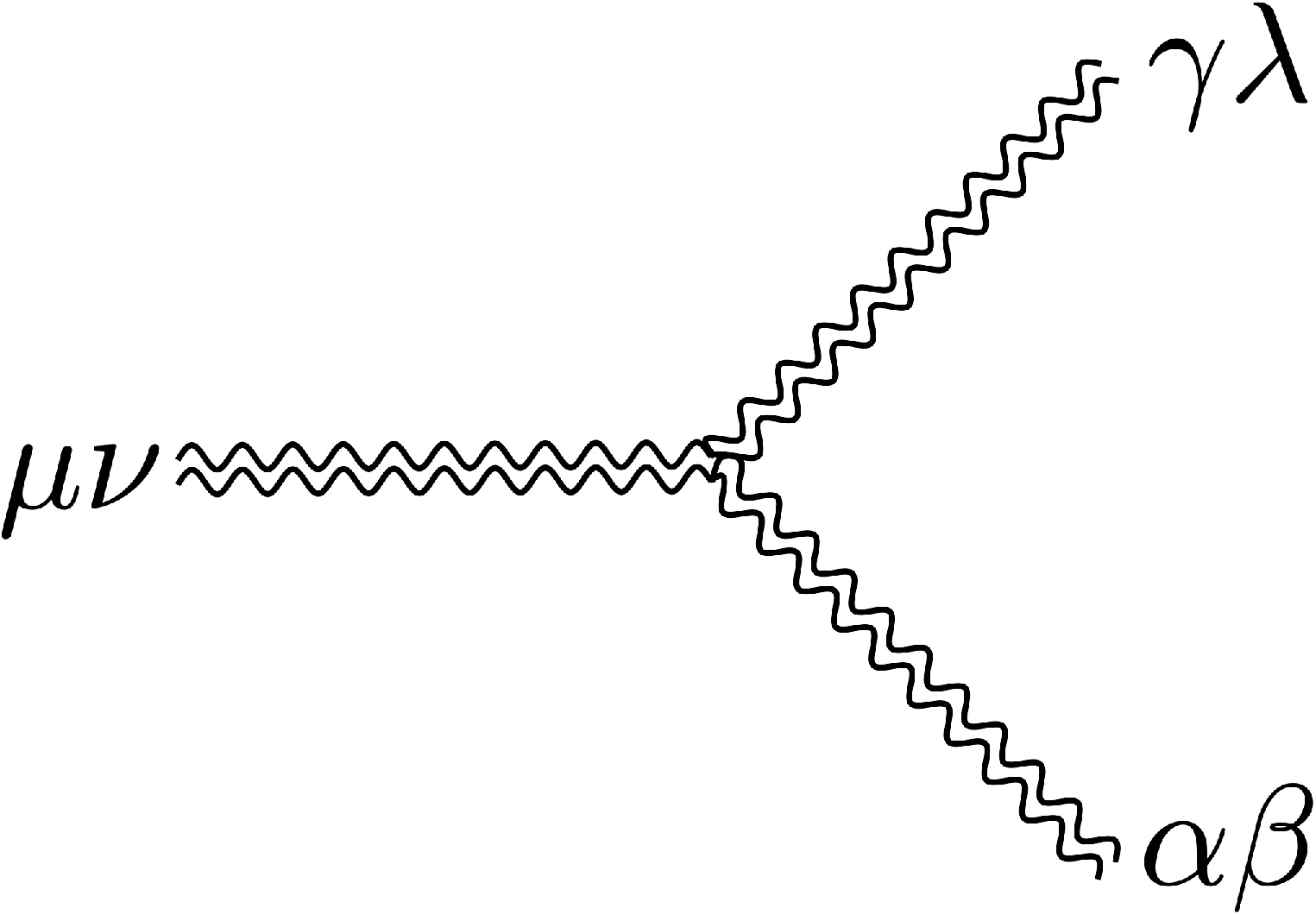}
    \hspace{5mm}
    \includegraphics[clip,width=3.6cm]{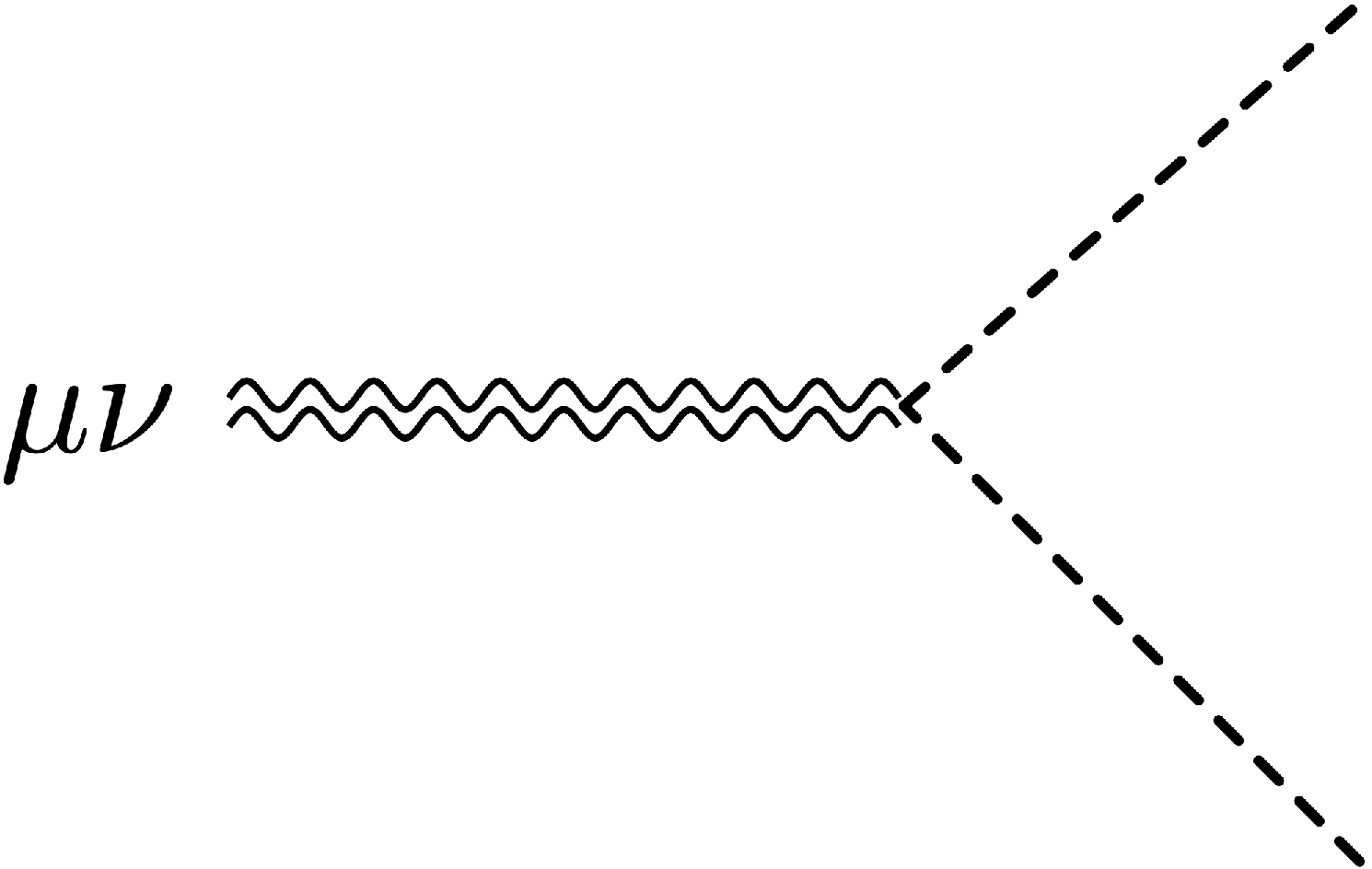}
    \hspace{5mm}
    \includegraphics[clip,width=3.0cm]{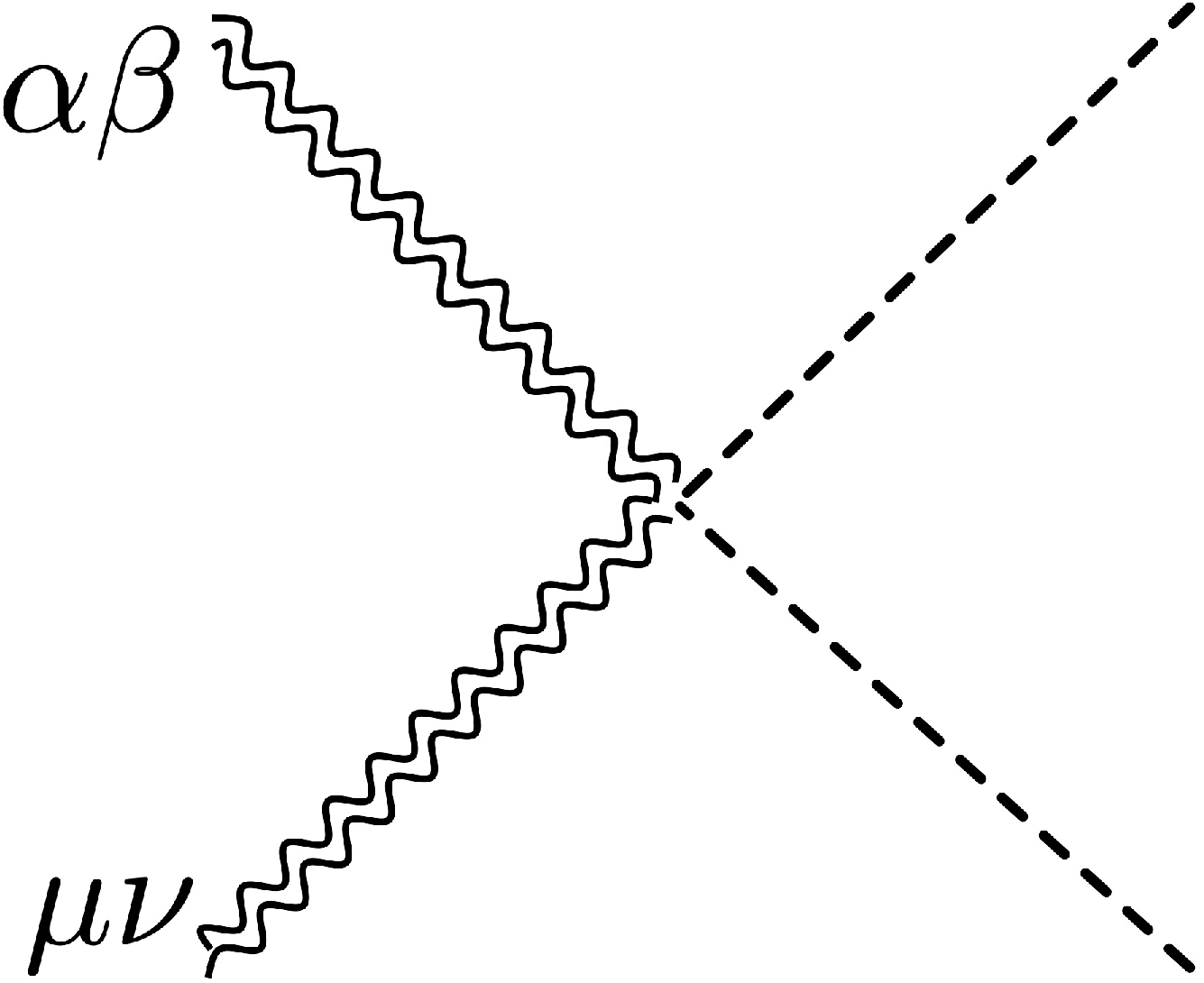}
  \caption{Graviton three-point (left), graviton-matter three-point (middle) and the graviton-matter four-point (right) vertex functions.}
    \label{Fig:ver}
  \end{center}
\end{figure}

We give Feynman rules.
$H_{\mu\nu}$ and $I_{\mu\nu}$ may be denoted by $h_{\mu\nu}$ collectively in Feynman diagrams,
since graviton field is expressed in terms of $h_{\mu\nu}$ in the action. 
Propagators and the vertices $h_{\mu\nu}h_{\alpha\beta}h_{\gamma\lambda}$, $h_{\mu\nu} \phi \phi $, $h_{\mu\nu} h_{\alpha\beta} \phi \phi $  are shown 
in Fig.\ref{Fig:pro} and \ref{Fig:ver}. 
These are the minimum requirements for computing $h_{\mu\nu}$-$\phi$ scattering at tree level. 
Feynman rules are obtained by expanding the action \eq{actiong} and \eq{actionm} in powers of $h_{\mu\nu}$ , as obtained in the forthcoming paper \cite{AII}. 
It is useful to define the transverse part $\h_{\mu\nu}$ of graviton $h_{\mu\nu}$, obeying $\p^\mu \h_{\mu\nu}=0$. 
Note that $\h_{\mu\nu}$ includes all on-shell states, that is, massless graviton $H_{\mu\nu}$ and massive graviton $I_{\mu\nu}$. 
We can write Feynman rules with $h_{\mu\nu}$ replaced by $\h_{\mu\nu}$.
This is because, in tree-level approximation which we will take, 
the calculation requires only degrees of freedom appearing as physical onshell modes, even for off-shell propagators.
The propagators for $\phi$ and $\h_{\mu\nu}$ are then given by 
\beqa
&&G_\phi =  \frac{-i}{p^2+m^2},\\
&&G_{\alpha\beta,\mu\nu}=\frac{2i}{\beta} \frac{1}{p^4 +m_I^2 p^2}P^{(2)}_{\alpha \beta, \mu \nu} 
+ \frac{i}{2(3\alpha+\beta)}\frac{1}{p^4 + m_S^2p^2} P^{(0)}_{\alpha \beta, \mu \nu}, \label{prog} \\
&&m_I^2 =- (\beta \kappa^2)^{-1}, \qquad m_S^2=(2\kappa^2(3\alpha+\beta))^{-1},
\eeqa
where $P^{(2)}_{\alpha \beta, \mu \nu}$ and $P^{(0)}_{\alpha \beta, \mu \nu}$ are the projections to the transverse-traceless and the transverse-trace part \cite{Abe:2017abx}. 
The first term in Eq.\eq{prog} shows the propagation of spin-2 degrees of freedom and may be written as
\beqa
\frac1{\beta}\frac{i}{ p^4+m_I^2  p^2} = - i \kappa^2 \left( \frac{1}{p^2}-\frac{1}{p^2+m_I^2}\right) ,
\eeqa
where $m_I$ is the massive graviton mass. 
The minus sign in the second term corresponds to $I_{\mu\nu}$ being a negative norm field. 
The second term in the right hand side of  Eq.\eq{prog} is also decomposed in the similar way. 
The mass of the scalar graviton is  $m_S$. 
The massless pole is gauge mode, and thus it does not appear as a physical onshell degrees of freedom.
The vertex functions are lengthy, and thus we do not show the explicit form here. 
They will be shown in the forthcoming paper\cite{AII}.
The difference between $H_{\mu\nu}$ and $I_{\mu\nu}$ appears on the external lines, 
that is, according to the initial and final gravitons being massless spin-2, massive spin-2 or scalar graviton, we operate the corresponding basis $e_{\mu \nu}^{(\sigma)}(p)$, $e_{\mu \nu}^{(\tau)}(p)$ or $\theta_{\mu\nu}(p)/ \sqrt{3}$, respectively, on the external graviton legs.

%
\section{Scattering amplitudes}
The purpose of this letter is to study the optical theorem \eq{OT} for $h_{\mu\nu}$-$\phi$
scattering in the lowest order of perturbation in $\kappa$. 
The invariant amplitude $\A$ is defined by 
\beqa
\langle \Phi |T | \Psi \rangle =\delta^4 (p_\Psi -p_\Phi) \A (\Psi \to \Phi) .
\eeqa
For $2\to2$ scattering $\A$ has the mass dimension $[\A]=0$.
 
We are concerned about the interplay between the massless and massive gravitons appearing as the intermediate states $\Phi$ in the optical theorem \eq{OT}. 
Hence we consider $h_{\mu\nu}$-$\phi$ scattering involving both massless (positive norm) and massive (negative norm) gravitons. 
The simplest such $2\to 2$ processes are:
\beqa  
&&
H^{(\sigma)}+ \phi \to H^{(\sigma')}+ \phi, \quad 
H^{(\sigma)}+ \phi \to I^{(\tau')}+ \phi, \quad 
I^{(\tau)}+ \phi \to I^{(\tau')}+ \phi, \nonumber \\ 
&&
H^{(\sigma)}+ \phi \to I^{(S)}+ \phi, \qquad 
I^{(\tau)}+ \phi \to I^{(S)}+ \phi, \qquad 
I^{(S)}+ \phi \to I^{(S)}+ \phi.
\label{pro}
\eeqa
We fix the scattering kinematics by 
\beqa
h \bigl(k_1,e_{1,\mu\nu}(k_1) \bigr) + \phi (k_2) \to h \bigl(k_3,e_{3,\alpha\beta}(k_3)\bigr) + \phi (k_4),
\eeqa
where $h$ indicates all modes of graviton $H^{(\sigma)}$, $I^{(\tau)}$ and $I^{(S)}$. 
$e_{i,\mu\nu} (k_i)$ stands for the corresponding bases $e^{(\sigma)}_{\mu\nu}(k_i)$, $e^{(\tau)}_{\mu\nu}(k_i)$ and 
$\theta_{\mu\nu} (k_i)/ \sqrt{3}$ ($i=1,3$).
We take the center of mass (CoM) frame, and set 
\beqa
&&\!\! k_{1,\mu} =\left(\sqrt{k^2 + m_1^2} ,k,0,0\right), \quad\  
k_{3,\mu} =\left(\sqrt{q^2 + m_3^2} ,q\cos\theta,q\sin\theta,0\right), \nonumber \\
&&\!\! k_{2,\mu} =\left(\sqrt{k^2 + m^2} ,-k,0,0\right), \quad
k_{4,\mu} =\left(\sqrt{q^2 + m^2} ,-q\cos\theta,-q\sin\theta,0\right), 
\eeqa
where $m_1^2, m_3^2 = 0$, $m_I^2$ or $m_S^2$ for massless, massive spin-2 or massive scalar graviton, respectively. 

To define bases $e^{(\sigma)}_{\mu\nu}$, $e^{(\tau)}_{\mu\nu}$ and $\theta_{\mu\nu}/\sqrt{3}$, 
we introduce longitudinal vector $l_{i,\mu}$ and transvers vectors $t_{i,\mu}$ and $u_{\mu}$ ($i=1,3$),
\beqa
&&l_{1,\mu} = m_1^{-1} \left( k,\sqrt{k^2 + m_1^2},0,0\right), \quad t_{1,\mu}= (0,0,1,0), \quad u_{\mu}= (0,0,0,1), 
\label{vbasis}
\\ 
&&l_{3,\mu} = m_3^{-1} \left( q,\sqrt{q^2 + m_3^2}\cos\theta,\sqrt{q^2 + m_3^2} \sin \theta,0\right), \quad
 t_{3,\mu}= (0,-\sin\theta,\cos\theta,0),  \nonumber
\eeqa
where $t_{i,\mu}$ $(i=1,3)$ is tangent to the spatial scattering plane but $u$ is normal to. 
The bases for graviton are expressed with these vectors,
\begin{eqnarray}
&&e_{i,\mu\nu}^{(0)} = \frac2{\sqrt {6}} l_{i,\mu}l_{i,\nu}-\frac1{\sqrt{6}} t_{i,\mu}t_{i,\nu}-\frac1{\sqrt{6}} u_{\mu}u_{\nu} ,\quad
e_{i,\mu\nu}^{(1,e)} = \frac1{\sqrt {2}} \left( l_{i,\mu}t_{i,\nu}+ t_{i,\mu}l_{i,\nu} \right), \nonumber \\
&&e_{i,\mu\nu}^{(1,o)} = \frac1{\sqrt {2}} \left( l_{i,\mu}u_{\nu}+ u_{\mu}l_{i,\nu} \right),\ 
e_{,i\mu\nu}^{(2,e)} = \frac1{\sqrt {2}} \left( t_{i,\mu}t_{i,\nu}- u_{\mu}u_{\nu} \right), 
\label{tbasis}
\\
&&e_{i,\mu\nu}^{(2,o)} = \frac1{\sqrt {2}} \left( t_{i,\mu}u_{\nu}+ u_{\mu}t_{i,\nu} \right), \quad
e_{i,\mu\nu}^{(s)} = \frac1{\sqrt {3}} \left( l_{i,\mu}l_{i,\nu}+t_{i,\mu}t_{i,\nu}+u_{\mu}u_{\nu} \right) = \frac1{\sqrt{3}} \theta_{i,\mu\nu} . \nonumber
\end{eqnarray}
Here, the numbers in the index show the helicity. 
For massless graviton basis $e^{(\sigma)}_{\mu\nu}$, these bases except helicity-2 are ill-defined, 
since the inverse of mass appears in $l_{i,\mu}$.
However, it is harmless,  
because massless graviton has only helicity-2 degrees of freedom. 
Indices $e$ and $o$ mean even and odd; 
in the bases with $e$ ($o$) index, all terms have even (odd) number of $u$. 
The scattering amplitude from even to odd (or vice versa) vanishes.

\begin{figure}[tb]
  \begin{center}
  \begin{tabular}{c}
      \begin{minipage}{0.22\hsize}
        \begin{center}
    \includegraphics[clip,width=2.3cm]{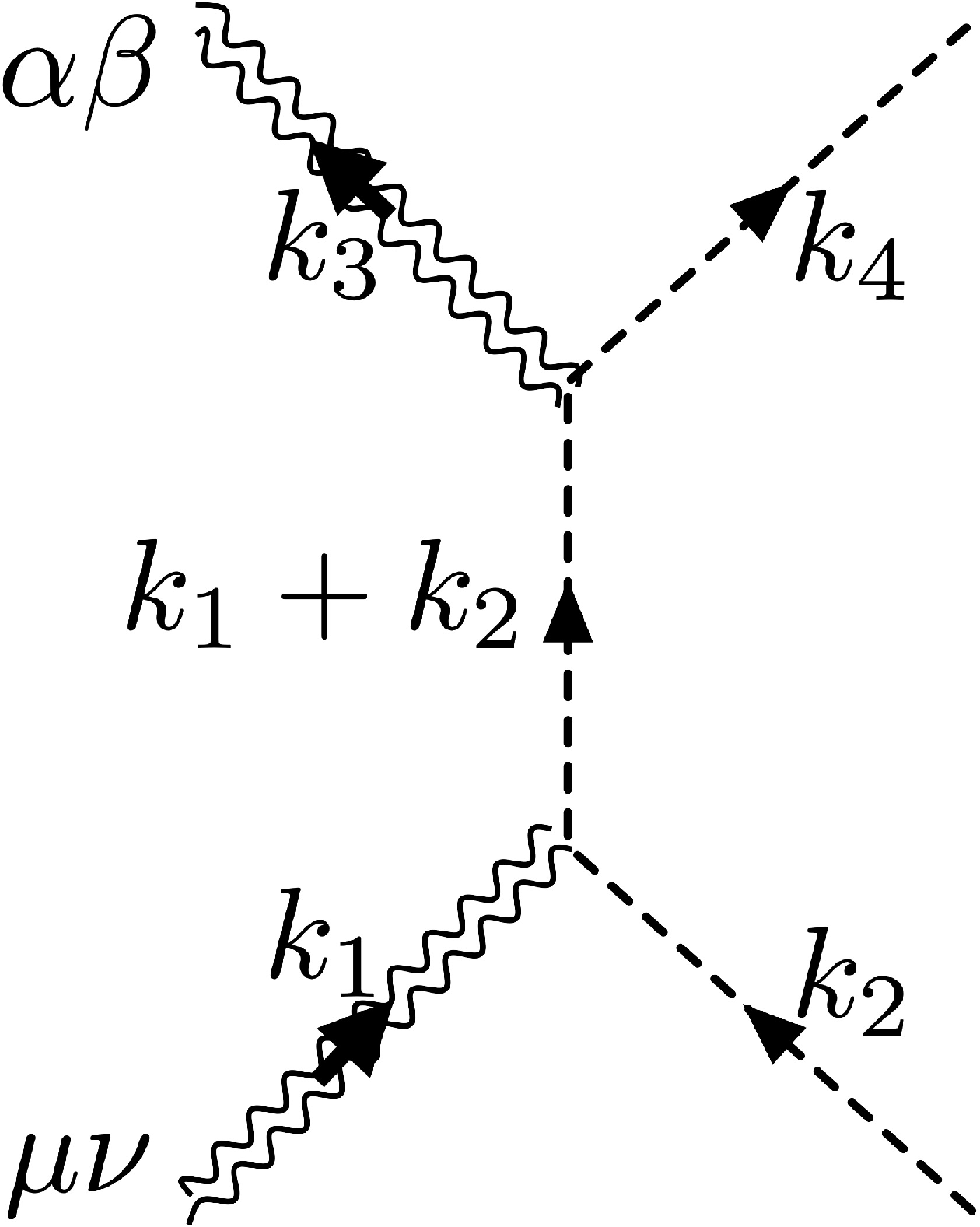}
        \end{center}
      \end{minipage}
      \begin{minipage}{0.22\hsize}
        \begin{center}  
    \includegraphics[clip,width=2.3cm]{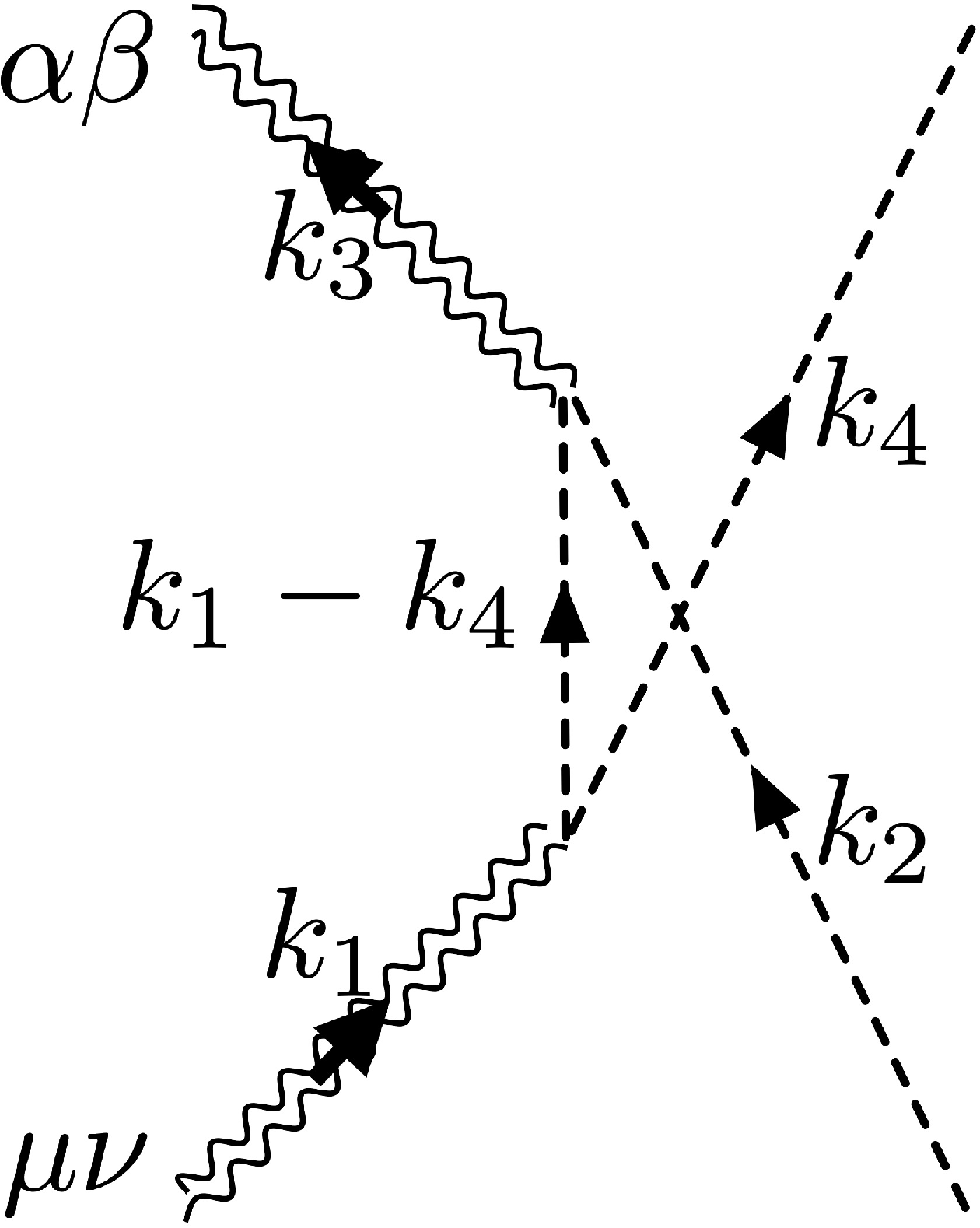}
        \end{center}
      \end{minipage}
      \begin{minipage}{0.32\hsize}
        \begin{center}  
    \includegraphics[clip,width=3.5cm]{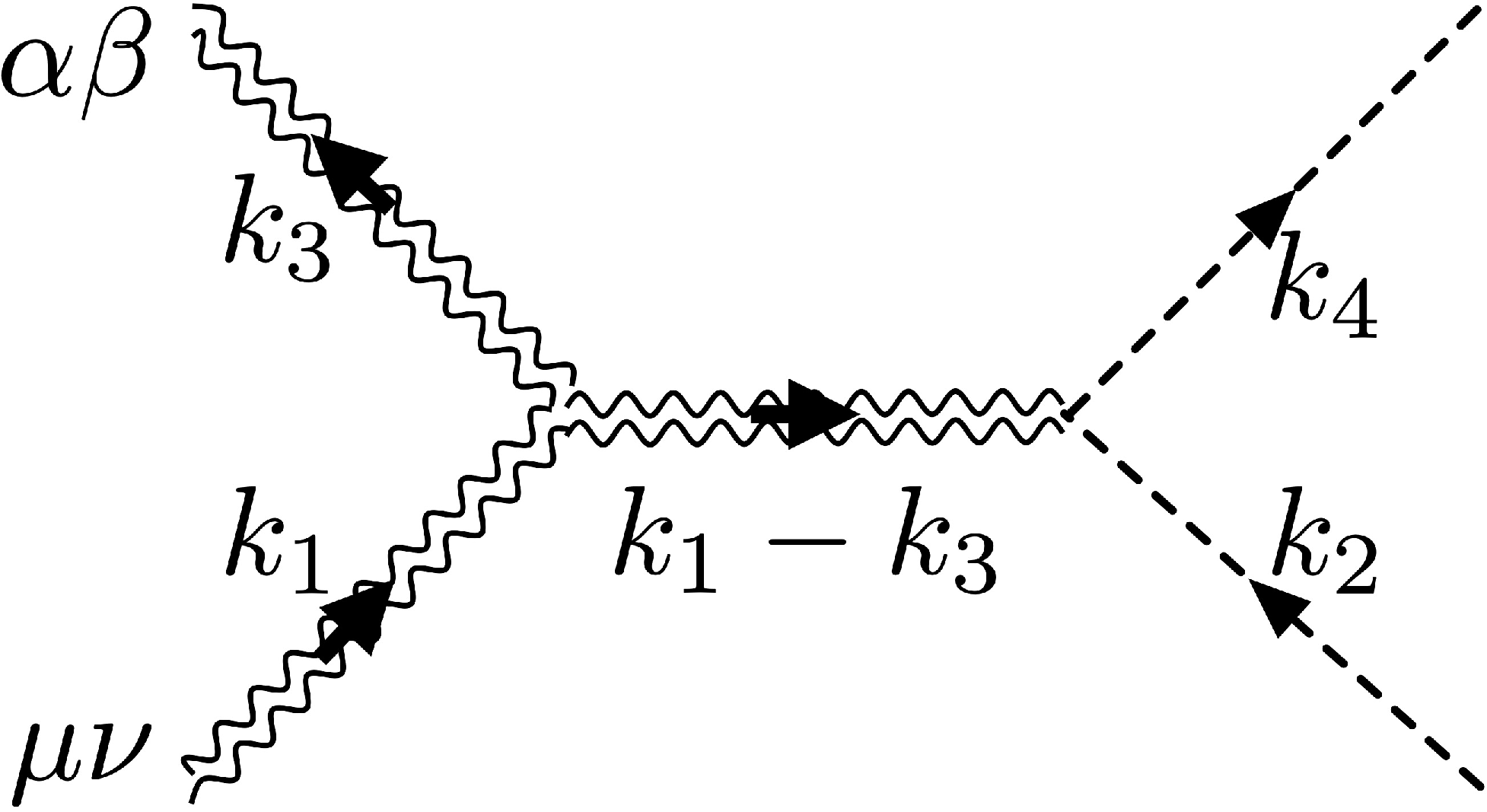}
        \end{center}
      \end{minipage}
      \begin{minipage}{0.25\hsize}
        \begin{center}  
    \includegraphics[clip,width=2.5cm]{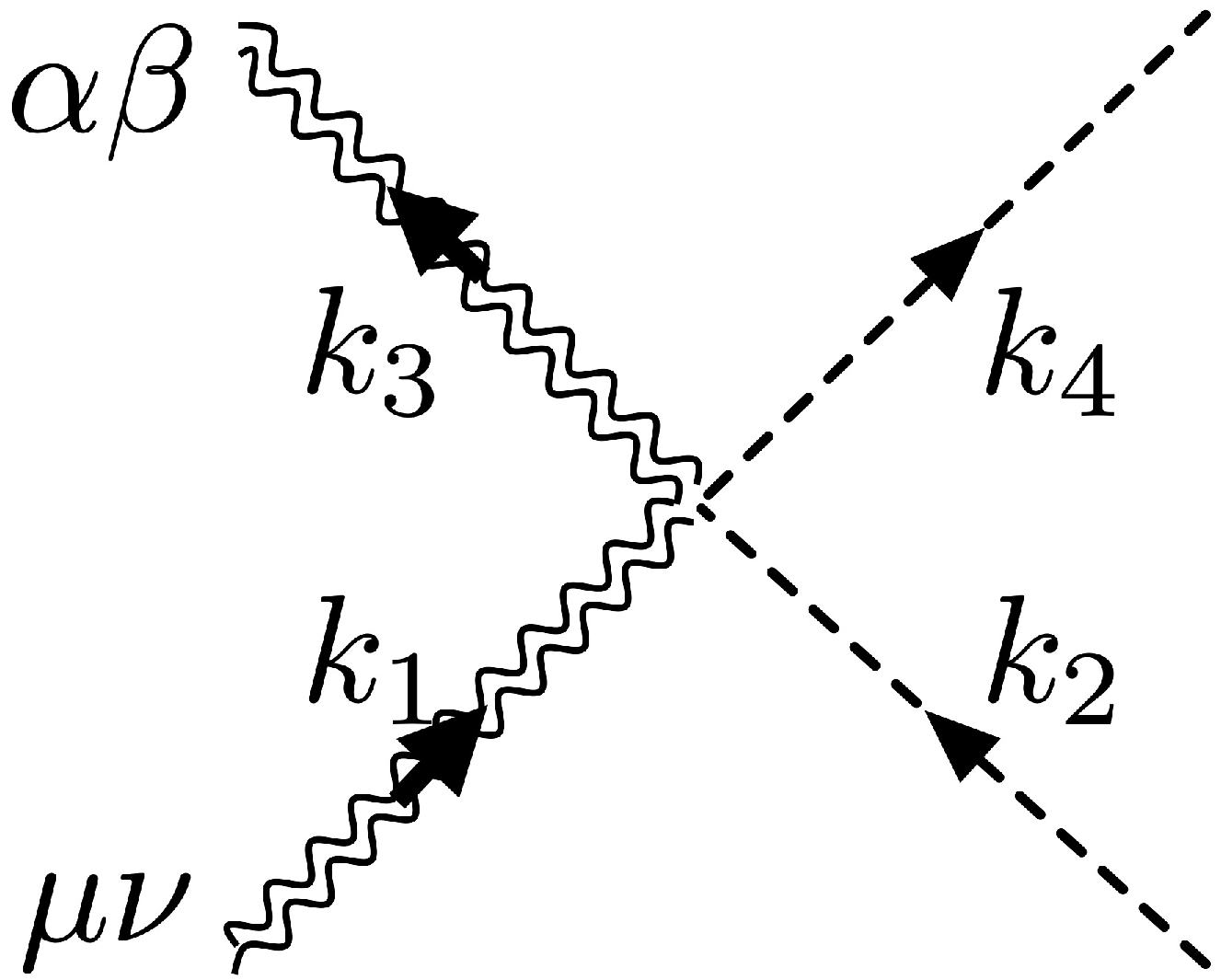}
        \end{center}
      \end{minipage}
\end{tabular}
    \caption{$s$-, $u$-, $t$-channels and contact diagrams (from left to right).}
    \label{Fig:cha}
  \end{center}
\end{figure}

We compute the amplitudes of \eq{pro} at tree level. 
Four types of graphs contribute to the scattering, contact term, $s$-channel and $u$-channel exchanges of $\phi$ propagator, $t$-channel exchange of $h_{\mu\nu}$ propagator (Fig.\ref{Fig:cha}), which are denoted by $\A_c$, $\A_s$, $\A_u$, $\A_t$, respectively.  
$\A_s$ and $\A_u$ are related by the crossing symmetry.
Because the difference among the amplitudes of \eq{pro} caused only by the on-shell bases operated to the external lines 1 and 3, 
these amplitudes are related to each other. 
Especially, the three amplitudes for helicity-2 mode on the first line of \eq{pro} and two amplitudes for helicity-2 mode of the first two on the second line of \eq{pro} are related, respectively, by the replacement $m_I^2 \leftrightarrow  m_H^2 =0$. 

The energy dependence of the scattering amplitudes $\A$ in high energy limit can be obtained as follows. 
The $s$- and $u$-channel exchange graphs have two $\phi^2 h_{\mu\nu}$ vertices, each of which gives the power $E^2$, 
and one scalar propagator with power $E^{-2}$. 
We obtain the power $E^2$ by combining the vertices and the propagator. 
As for the $t$-channel, the power $E^2$ is obtained from the $\phi^2 h_{\mu\nu}$ with the power $E^2$, 
$\phi h_{\mu\nu} h_{\alpha\beta}$ with the power  $E^4$ and propagator with the power  $E^{-4}$.
The contact diagram gets the power  $E^2$ from the $\phi^2 h_{\mu\nu} h_{\alpha\beta}$ vertex. 
Since all diagrams have the power  $E^2$, their sum is also estimated as the power  $E^2$. 
There is another source in the scattering amplitude;
it picks up $E$ dependence of the helicity bases of the initial and final states. 
As seen in Eqs.(\ref{vbasis}) and (\ref{tbasis}),
the bases have the power  $E^{2-a}$, where $a$ is the helicity of the basis. 
Therefore, the amplitude $\A$ is estimated as
\beqa
\A = \beta E^\alpha, \qquad \alpha={6-a_i-a_f}, 
\label{naivees}
\eeqa
where $a_i$ and $a_f$ are the helicities of the initial and final gravitons respectively, 
unless the nontrivial cancelation occurs.

Let us see the scattering amplitudes in more details.
The computation requires many pages and is relegated to the forthcoming paper\cite{AII}. 
Here we only give the amplitudes in the high energy (HE) limit $E\to\infty$ (which is equivalent to $k\to\infty$), where the CoM energy $E$ is $E=k_{10} + k_{20}=2k + (m_1^2+m^2)/(2k)$.
The scattering amplitude $\A$ is the sum of the four terms, $\A =\A_c + \A_s + \A_u + \A_t$ .
\beqa
&&
\A\left(h^{(2,o)}+\phi \to h^{(2,o)}+\phi\right) = \A\left(h^{(2,e)}+\phi \to h^{(2,e)}+\phi\right) 
\nonumber \\
&& \hspace{45mm}
=
-\kappa^2 k^2\frac{1+\cos \theta }{1-\cos \theta}
+\cO\left( k^0 \right),
\label{AtoA1}
\\
&&
\A\left(h^{(2,o)}+\phi \to I^{(1,o)}+\phi\right) = -\A\left(h^{(2,e)}+\phi \to I^{(1,e)}+\phi\right) 
\nonumber \\
&& \hspace{45mm}
=
\kappa^2\frac{m_I k  \sin \theta }{2(1-\cos \theta)}
+\cO\left( k^{-1} \right) ,
\\
&&
\A\left(I^{(1,o)}+\phi \to I^{(1,o)}+\phi\right) = \A\left(I^{(1,e)}+\phi \to I^{(1,e)}+\phi\right) 
\nonumber \\
&&\hspace{30mm}
=-\frac{\kappa^2 m_I^2}{8} \frac{2^2 +(1+\cos\theta)^2 +(1-\cos\theta)^2 }{(1-\cos \theta)^2}+ 
\cO\left( k^{-2} \right),
\\
&&
\A\left(h^{(2,e)}+\phi \to I^{(0,e)}+\phi\right) = \cO(k^0) ,
\\
&&
\A\left(I^{(1,e)}+\phi \to I^{(0,e)}+\phi\right) =  \cO(k^{-1}) ,
\\
&&
\A\left(I^{(0,e)}+\phi \to I^{(0,e)}+\phi\right) = -
\frac{\kappa^2}{2} \frac{m_I^2(1+\cos \theta)}{(1-\cos \theta)^2}  -\frac{\kappa^2}{8} (m_I^2 +2m^2)   + \cO(k^{-2}) ,
\nonumber\\
&&\\
&&
\A\left(h^{(2,e)}+\phi \to I^{(S)}+\phi\right)  
\nonumber\\
&&\hspace{20mm}
=-
\frac{\sqrt{6}\kappa^2}{48} (m_S^2+2m^2)(3+\cos^2 \theta) -\sqrt{6} \xi\kappa^2 m_S^2 + \cO(k^{-2}),
\\
&&
\A\left(I^{(1,e)}+\phi \to I^{(S)}+\phi\right)  =  \cO(k^{-1}),
\\
&&
\A\left(I^{(0,e)}+\phi \to I^{(S)}+\phi\right)  =  \cO(k^{0}) ,
\\
&&
\A\left(I^{(S)}+\phi \to I^{(S)}+\phi\right)  
= \frac{\kappa^2}{24m_S^2} (4m_S^4 -8m_S^2m^2+9m^4) 
\nonumber \\
&& \hspace{35mm}
-\frac{1}{72 \kappa^2 \beta} \left( 1 -\frac{12(1-\cos\theta)}{(1+\cos\theta)^2} \right) 
+8\xi \kappa^2 m_S^2 
+  \cO(k^{-2}),
\label{AtoA2}
\eeqa
and the others are zero, where $h$ is $H$ or $I$.

The HE behavior of these amplitudes may also be written as 
\beqa
&&  \A(h^{(a,b)} +\phi \to h^{(a',b')} + \phi)  \sim \beta_{(a,b),(a'b')} E^{\alpha_{aa'}} ,  \\
&&  \A(h^{(a,b)} +\phi \to I^{(S)} + \phi)  \sim \beta_{(a,b)} E^{\alpha_{aS}} ,  \\
&&  \A(I^{(S)} +\phi \to I^{(S)} + \phi)  \sim \beta E^{\alpha_{SS}} ,  
\eeqa
where $a,a'=2,1,0$ and $b,b' = o,e$. 
Equations \eq{AtoA1}-\eq{AtoA2} imply
\beqa
&&
\alpha_{22} = 2, \qquad \alpha_{21} = 1, \qquad \alpha_{20} = 0,  \qquad \alpha_{11} = 0, \qquad \alpha_{10} \le -1, 
\nonumber \\
&&
\alpha_{00} = 0, \qquad \alpha_{2S} = 0, \qquad \alpha_{1S} \le -1, \qquad \alpha_{0S} \le 0, \qquad \alpha_{SS} = 0.
\label{ED}
\eeqa 
The HE behavior of the elastic $2\to2$ amplitudes for the massless graviton-matter scattering is  
\beqa
\A\left( H^{(2,e)} +\phi \to H^{(2,e)} +\phi \right)\sim E^2.
\eeqa
It is the same as that in Einstein gravity \cite{DeWitt3,BG}, and tree unitarity \eq{Aineq} is apparently violated. 
Hence, it has a lack of tree unitarity at high energy.
The immediate question is whether the optical theorem \eq{OT} is still obeyed or not. We will study this question using the the amplitudes obtained above.

Note that the energy dependence in Eq.(\ref{ED}) is different from the naive estimate in Eq.(\ref{naivees}). 
This is because the nontrivial cancelations occur among different channels due to the symmetry. 
The cancelations are required for the theory to have the perturbative $S$-matrix unitarity.
A similar cancelation was seen in the Weinberg-Salam theory\cite{Lee:1977eg}, 
where the authors have shown that the theory with only massive vectors does not satisfy tree unitarity
but, introducing symmetry by Higgs field, tree unitarity holds. 
Similarly, if we introduce something violating symmetry, 
for instance considering a positive-norm  massive graviton and/or interactions violating the symmetry, the power of $E$ of the scattering amplitude becomes higher, 
and then the perturbative $S$-matrix unitarity is broken.  
This is a good example of showing the equivalence between perturbative $S$-matrix unitarity and renormalizability.
(Even if the dimension of coupling constant is zero, the theory is NOT renormalizable\cite{FIIK1,FIIK2}.)

%

\begin{figure}[tb]
  \begin{center}
  \begin{tabular}{c}
      \begin{minipage}{0.24\hsize}
        \begin{center}
    \includegraphics[clip,width=2.4cm]{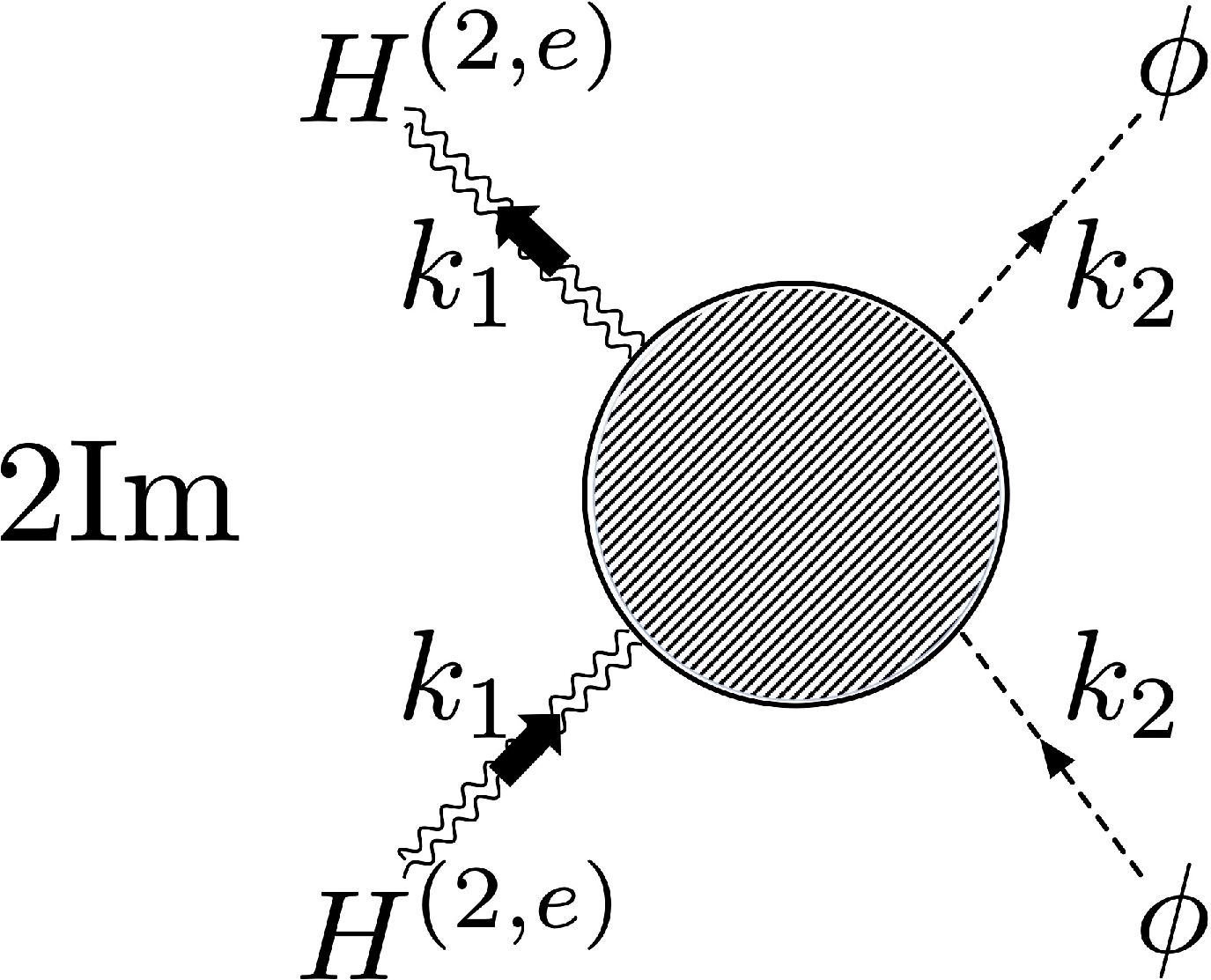}
        \end{center}
      \end{minipage}
      \begin{minipage}{0.24\hsize}
        \begin{center}  
    \includegraphics[clip,width=3.0cm]{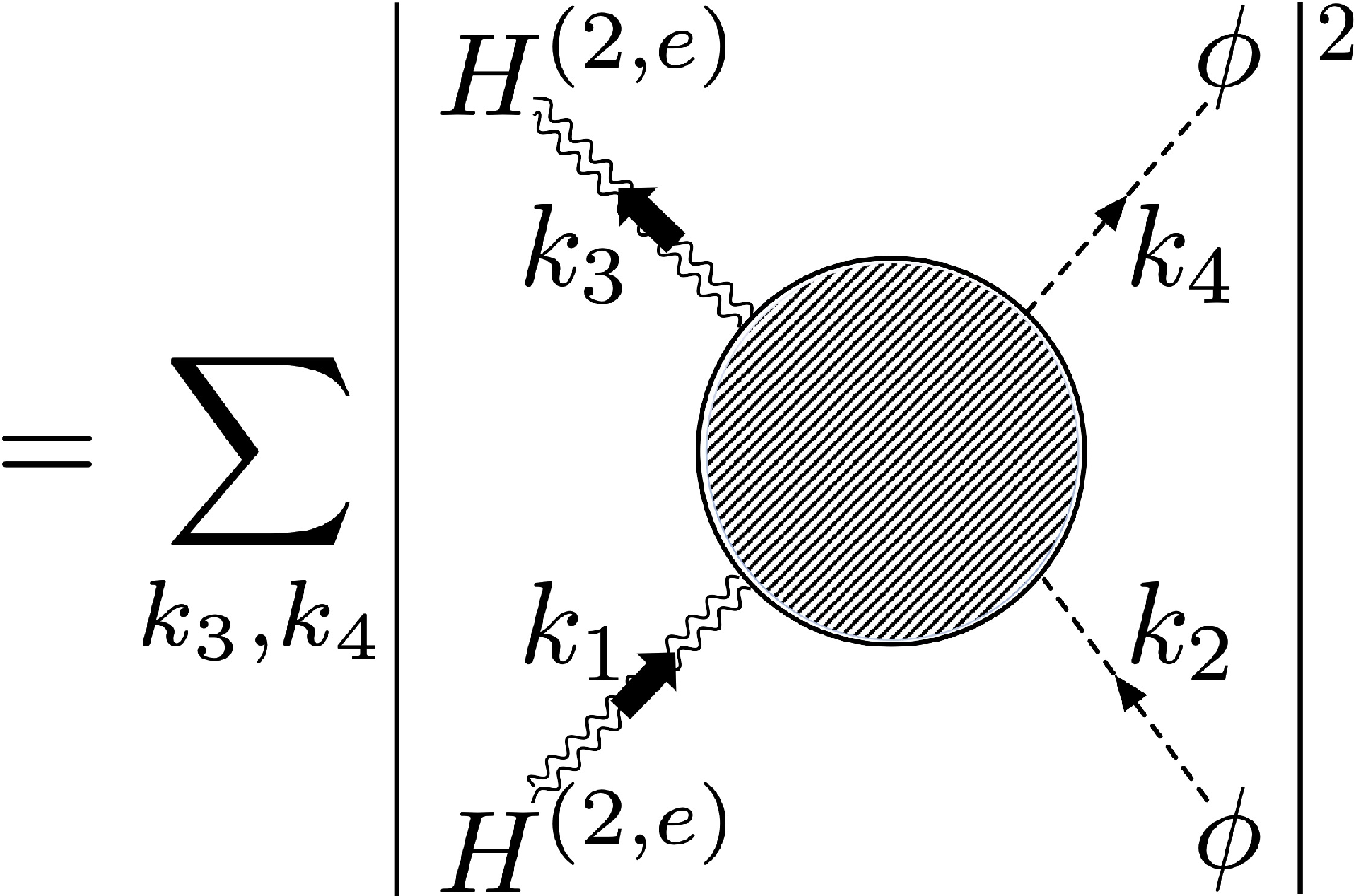}
        \end{center}
      \end{minipage}
      \begin{minipage}{0.24\hsize}
        \begin{center}  
    \includegraphics[clip,width=3.2cm]{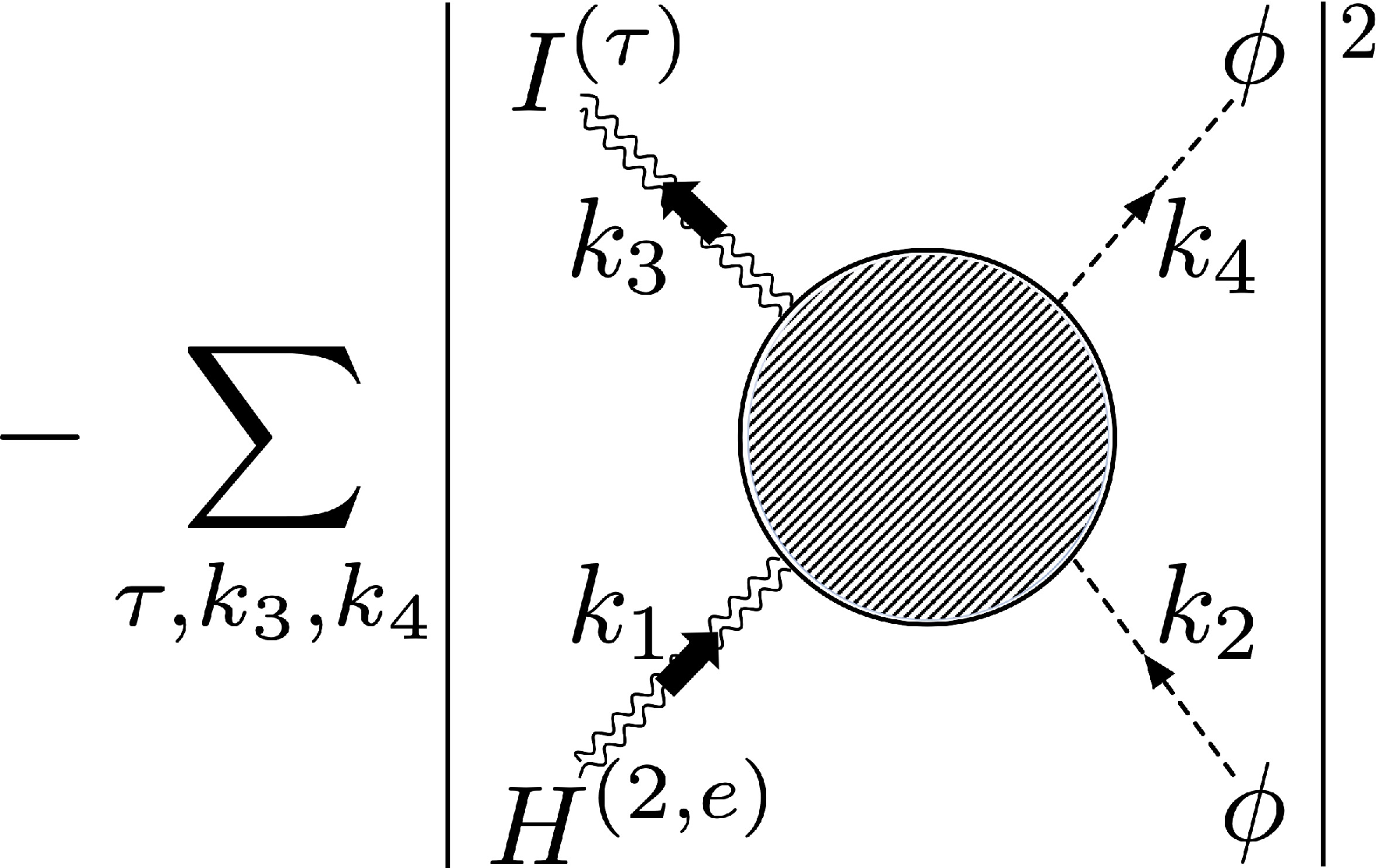}
        \end{center}
      \end{minipage}
      \begin{minipage}{0.24\hsize}
        \begin{center}  
    \includegraphics[clip,width=3.0cm]{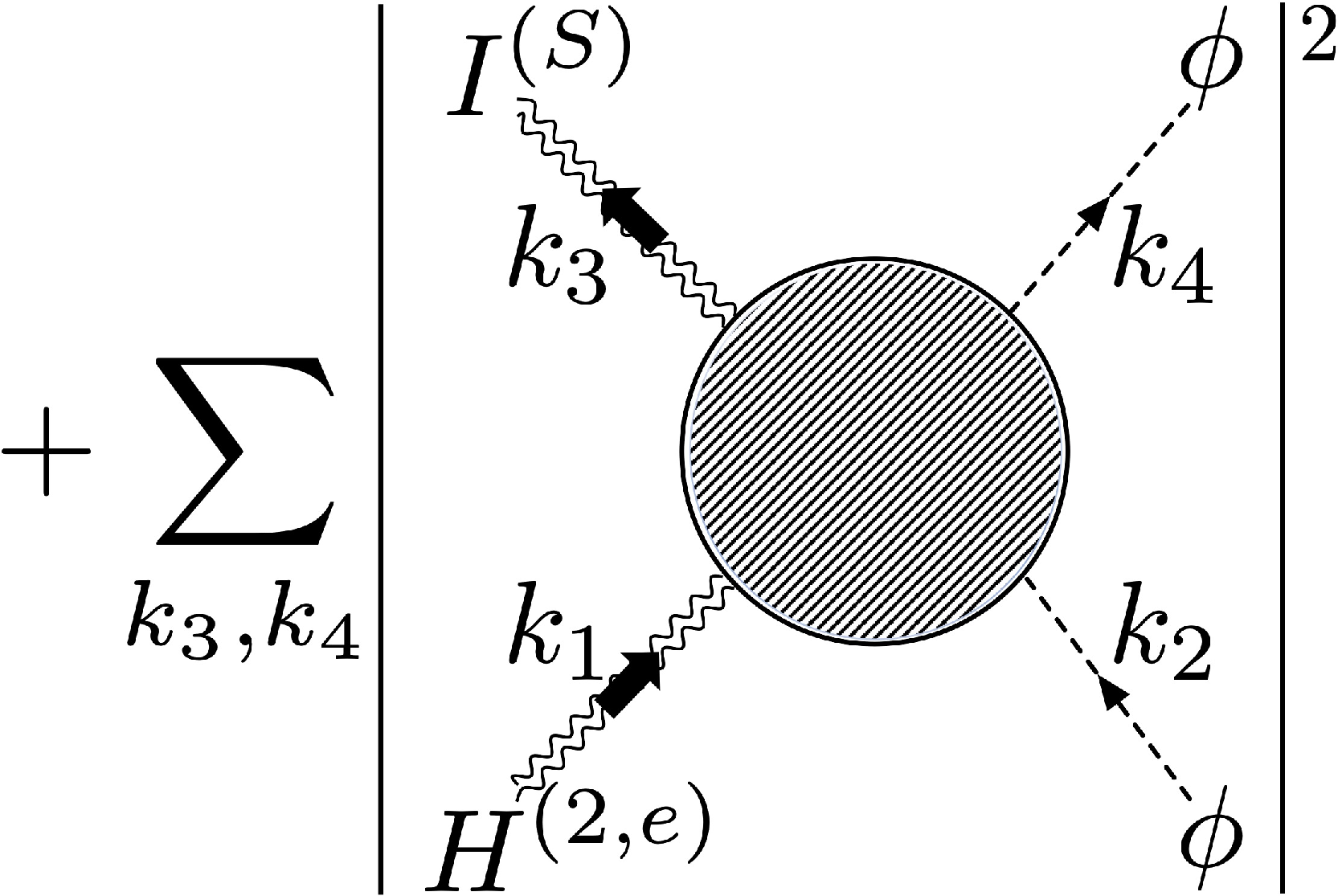}
        \end{center}
      \end{minipage}
\end{tabular}
    \caption{The optical theorem for $H^{(2,e)}$-$\phi$ scattering in the two-particle approximation.}
    \label{Fig:OT}
  \end{center}
\end{figure}

\section{S-matrix unitarity }
Finally we will show the mechanism how the condition $\langle\Psi|S^\dagger S|\Psi\rangle=1$ is met due to the cancellation of the two contributions, one from positive norm graviton $H_{\mu\nu}$ and the other from negative norm graviton $I_{\mu\nu}$ in the intermediate sum in the optical theorem \eq{OT}. 
We apply Eq.\eq{OT} to the elastic scattering $h^{(\sigma)} +\phi \to  h^{(\sigma)} +\phi$, {\it i.e.} $\Psi = h^{(\sigma)} +\phi$ as depicted in Fig \ref{Fig:OT}. 
We take the two-particle approximation of the intermediate states $\Phi$, which is the lowest order in the perturbation in $\kappa$, 
hence $\Phi =\{H_{\mu\nu} +\phi\}$ and $\{I_{\mu\nu} +\phi\}$ in Eq.\eq{OT}.

We demonstrate 
Eq.\eq{SOTineq} 
by $H^{(2,e)} +\phi\to H^{(2,e)} +\phi$. In the present two-particle approximation for $\Phi$, 
noting $\epsilon_{H^{(\sigma)} +\phi}=+1$, $\epsilon_{I^{(\tau)} +\phi}=-1$ and $\epsilon_{I^{(S)} +\phi}=+1$, we have%
\footnote{
To be accurate, the sum over the intermediate states on the right 
side includes the integration with respect to three-dimensional momenta, 
but it is not shown explicitly. Furthermore, the momentum-dependent normalization factor is required. 
In $2\to2$ scattering in four-dimensional spacetime, such dependence accidentally cancel to each other. 
Generic cases have been discussed\cite{FIIK2}.
Moreover, we use the fact that the amplitudes involving both even and odd modes vanish.}
\beqa
&&\left| \A \left(H^{(2,e)} +\phi\to H^{(2,e)} +\phi \right)\right| \ge 
 \left| \A \left( H^{(2,e)} +\phi\to H^{(2,e)} +\phi\right)\right| ^2 \nonumber \\
&& \hspace{5mm}
-\sum_{\tau} \left| \A \left( H^{(2,e)} +\phi\to I^{(\tau)} +\phi\right)\right| ^2
+ \left| \A \left( H^{(2,e)} +\phi\to I^{(S)} +\phi\right)\right| ^2 .
\label{HOT}
\eeqa
where $(\tau) = (2,e),(1,e),(0)$ and we ignore the unimportant numerical factors.
For the optical theorem \eq{OT} to be satisfied in HE limit, 
the $k$ dependence of the left hand side of inequality of Eq.\eq{HOT} cannot be weaker than that of the right hand side. 
We evaluate the both sides using the tree amplitudes obtained in sec.3. 
The left hand side of Eq.\eq{HOT} is bounded as
\beqa
\left|  \A\left(H^{(2,e)}+\phi\to H^{(2,e)}+\phi\right) \right|
= \kappa^2 k^2 \frac{1+\cos\theta}{1-\cos\theta} + \cO\left(k^0\right).
\eeqa
We explicitly write down the right side of inequality \eq{HOT},
\beqa
&&\left|\A\left(H^{(2,e)}+\phi \to H^{(2,e)}+\phi\right)\right|^2
-\left|\A\left(H^{(2,e)}+\phi \to I^{(2,e)}+\phi\right)\right|^2
\nonumber \\ && \hspace{10mm}
-\left|\A\left(H^{(2,e)}+\phi \to I^{(1,e)}+\phi\right)\right|^2
-\left|\A\left(H^{(2,e)}+\phi \to I^{(0,e)}+\phi\right)\right|^2
\nonumber \\ && \hspace{10mm}
+\left|\A\left(H^{(2,e)}+\phi \to I^{(S)}+\phi\right)\right|^2 .
\label{ex}
\eeqa
Since Eq.\eq{AtoA1} shows that the first term has $k^4$ dependence in the leading order, inequality of Eq.\eq{HOT} seems to be violated in the first glance.
However, if we look at the first two terms of the right hand side of Eq.\eq{ex}, their leading order $k^4$ dependence cancel to each other, 
and there remains the next leading $k^2$ term.
Since the other terms are $\cO(k^2)$,  Eq.\eq{ex} is indeed $\cO(k^2)$. 
Hence, the $k$ dependence is the same in the both sides of inequality of Eq.\eq{HOT}, 
thus satisfying the necessary condition for the $S$-matrix unitarity. 

For other elastic $2\to2$ amplitudes, Eqs.\eq{AtoA1} through \eq{AtoA2} show us that the similar inequality is satisfied due to the same cancelation between $H^{(2,b)}$ and $I^{(2,b)}$ ($b=o,e$) appearing in the sum of $\sigma$ and $\tau$, namely $\Phi$ in the intermediate sum of Eq.\eq{OT}. 
Hence $S$-matrix unitarity holds in  all matter-graviton scattering. 
Note again that $S$-matrix unitarity holds due to the symmetry. 
Without symmetry, the power of scattering amplitude behaves as Eq.(\ref{naivees}), 
and we can see by the similar analysis that the perturbative $S$-matrix unitarity is violated then.

%

\section{Discussion}
We suggested in our previous work\cite{Abe:2018rwb} that $S$-matrix unitarity can be a guideline to renormalizability, 
and we have shown in this letter that it is true in gravitational theory. 
Hence, in $R_{\mu\nu}^2$ gravity, which is a renormalizable theory, the perturbative $S$-matrix unitarity is satisfied due to the symmetry. 
On the other hand, if something violating the symmetry is introduced, the perturvative $S$-matrix unitarity does not hold. 
It is consistent with the fact that such a theory is not renormalizable\cite{FIIK1,FIIK2}.
If a kinetic term in the action is degenerate, such as that in gauge theory, 
because the power counting theorem does not work, the proof of renormalizability is hard. 
Renormalizability of the quadratic gravity was proved by Stelle \cite{Stelle} in BRST method, 
where additional degrees of freedom, BRST ghosts and gauge modes, 
need to be introduced. 
On the other hand, the discussion of $S$-matrix unitarity done here is at the tree level, and hence it does not involve unphysical modes. 
That is to say, the discussion of renormalizability often involves unphysical degrees of freedom, 
but that of $S$-matrix unitarity does not.
The result is an evidence of our conjecture that the conditions for $S$-matrix unitarity at tree level and renormalizability are identical. 
Although our conjecture is not proved completely yet, 
it is useful to study renormalizability of various theories; 
the on-shell amplitude at tree level would show renormalizability in full order.

Two-graviton scattering in the quadratic gravity is expected to satisfy $S$-matrix unitarity too. 
We leave the project to demonstrate it in future work.

Interesting extension of  the quadratic gravity is discussed by Modesto and his collaborators~\cite{Modesto:2011kw,Modesto:2014lga,Briscese:2018oyx}, where the infinite number of derivatives are introduced.
Due to the infiniteness of the number of derivatives, 
the theory does not suffer from negative norms by the higher order derivatives, 
but it is renormalizable. 
It would be interesting to see what happens in this theory.

\section*{Acknowledgements}
The authors would like to thank M.~Tanabashi for reading the manuscript and S.~Odintsov for letting us know the role of $\phi^2 R$ term in action. 
We owe to N.~Ohta for a valuable discussion of the negative 
norm field in connection with renormalizability of the $R^2_{\mu\nu}$ gravity.
Y.\,A. is supported by JSPS Grants-in-Aid for Early-Career Scientists (No.\,19K14719).
K.\,I. is supported by JSPS Grants-in-Aid for Scientific Research (B) (20H01902) and for Scientific Research (A)(No.\,17H01091).

\end{document}